\documentclass[prb,superscriptaddress,twocolumn,amsmath,amssymb,nofootinbib]{revtex4-2}

\usepackage{hyperref}
\usepackage{graphicx}
\usepackage{natbib}
\usepackage{physics}
\usepackage{braket}
\usepackage{caption}
\usepackage{subcaption}

\usepackage[dvipsnames]{xcolor}

\renewcommand{\rho}{\varrho}

\newcommand\be{\begin{equation}}
\newcommand\ee{\end{equation}}
\newcommand\la{\langle}
\newcommand{\ra}{\rangle}

\begin{document}

\title{Super-phenomena in arbitrary quantum observables}

\author{Andrew N. Jordan}
\affiliation{The Kennedy Chair in Physics, Chapman University, Orange, CA 92866, USA}
\affiliation{Institute for Quantum Studies, Chapman University, Orange, CA 92866, USA}
\author{Yakir Aharonov} 
\affiliation{Institute for Quantum Studies, Chapman University, Orange, CA 92866, USA}
\author{Daniele C. Struppa}
\affiliation{The Donald Bren Distinguished Presidential Chair in Mathematics, Chapman University, Orange, USA}
\affiliation{Institute for Quantum Studies, Chapman University, Orange, CA 92866, USA}
\author{Fabrizio Colombo}
\affiliation{Politecnico di Milano, Dipartimento di Matematica, Via E. Bonardi, 9 20133 Milano, Italy}
\author{Irene Sabadini}
\affiliation{Politecnico di Milano, Dipartimento di Matematica, Via E. Bonardi, 9 20133 Milano, Italy}
\author{Tomer Shushi}
\affiliation{Department of Business Administration, Guilford Glazer Faculty of Business and Management, Ben-Gurion University of the Negev, Beer-Sheva, Israel}
\author{Jeff Tollaksen}
\affiliation{Institute for Quantum Studies, Chapman University, Orange, CA 92866, USA}
\author{John C. Howell}
\affiliation{Institute for Quantum Studies, Chapman University, Orange, CA 92866, USA}
\author{A. Nick Vamivakas}
\affiliation{Institute of Optics, University of Rochester, Rochester, NY 14627, USA}

\date{\today}

\begin{abstract}
Superoscillations occur when a globally band-limited function locally oscillates faster than its highest Fourier coefficient. We generalize this effect to arbitrary quantum mechanical operators as a weak value, where the preselected state is a superposition of eigenstates of the operator with eigenvalues bounded to a range, and the postselection state is a local position.  Superbehavior of this operator occurs whenever the operator's weak value exceeds its eigenvalue bound.  We give illustrative examples of this effect for total angular momentum and energy.  In the later case, we demonstrate a sequence of harmonic oscillator potentials where a finite energy state converges everywhere on the real line, using only bounded superpositions of states whose asymptotic energy vanishes - ``energy out of nothing''. 
This limit requires postselecting the particle in a region whose size diverges in the considered limit. We further show that superenergy behavior implies that the state superoscillates in time with a rate given by the superenergy divided by the reduced Planck's constant.  This example demonstrates the possibility of mimicking a high-energy state with coherent superpositions of nearly zero-energy states for as wide a spatial region as desired.  We provide numerical evidence of these features to further bolster and elucidate our claims.
\end{abstract}

\maketitle
The subject of superoscillations has attracted great interest in a variety of communities.  Mathematically, issues of the behaviour and  convergence of superoscillatory functions of band-limited functions have been studied in great detail \cite{berry2006evolution,aharonov2017mathematics}.  Physically, their connections with optical phenomena and quantum mechanics continue to be a fruitful research enterprise \cite{berry2019roadmap,kempf2018four}.  Applications of superoscillations have recently begun, with the realization that this phenonemon enables superresolution beyond the Rayleigh criterion of optical imaging system, relying on the sub-wavelength structures of the point spread function \cite{huang2007optical,huang2009super,rogers2012super,zheludev2008diffraction,greenfield2013experimental,makris2011superoscillatory}.  Superoscillations began with the investigation of Yakir Aharonov and colleagues into the physics of weak measurement \cite{aharonov1990can,aharonov1990properties,berry1994faster}, and the current investigation is taking this line of research full circle, to generalize this notion to arbitrary observables in quantum mechanics.  While we focus here on a quantum treatment, the effects we describe can be applied to differential equations and operators more generally, and can naturally be extended to other wave equations, such as optical phenomena. 

The present work generalizes the notion of superoscillation by noticing that it may be viewed a special type of weak value of the momentum operator.  Consequently, generalizations of the concept can be had by replacing the momentum operator by an arbitrary operator.  By preselecting on a state of a superposition of bounded eigenvalues of the operator of interest, and postselecting on position, the suitable generalized notion of superoscillation is obtained.  Whenever the weak value exceeds its eigenvalue range for some position, we define this to be a point of superbehavior of the operator.  Examples of this phenomena are given in the case of total angular momentum and energy.

An outstanding issue in the field is the tradeoff between the range of superoscillation and the intensity (or power) of the function.  Recent work has proven that mathematical limits exist where a superoscillating sequence converges to a function that can superoscillate everywhere in its domain, despite remaining band-limited at every value of the sequence \cite{aharonov2011some,aharonov2021new}.  This motivates the question of whether this feature can also be extended to other examples.   We will answer this question in the affirmative, and explicitly construct an example where a finite energy state can be created out of asymptotically zero-energy states.  This situation also demands that the wavefunction superoscillate in time with a rate given by the same superenergy divided by the reduced Planck's constant.

We organize the paper as follows:  In Sec.~\ref{sec1}, we recast the phenomenon of superoscillations as a weak value of momentum in a quantum mechanical context.
In Sec.~\ref{sec:gen}, we generalize this phenomenon of superoscillations to any observable in quantum physics.  In Sec.~\ref{sec-ilex} the illustrative examples of total angular momentum and energy are given.  In Sec.~\ref{energy-nothing} a sequence of potentials are given such that in the limit as the iterator goes to infinity, a finite energy state is seemingly created out of zero energy ingredients.  In Sec.~\ref{sec-qmim}, further insight is given by calculating the energy two ways - the first is the weighted energy in the spectral basis, which indeed goes to zero.  However, if the expected energy is calculated in a finite spatial domain - whose width diverges as the sequence limits to infinity - the expected energy is indeed finite.  This situation corresponds to postselecting the particle within this spatial region.  In Sec.~\ref{sec-sotime} we demonstrate that a necessary consequence of superbehavior of an operator is the superoscillation of its generating variable.  We illustrate this fact with superoscillations in time for the rigid rotor (demonstrating super total angular momentum) and for the harmonic oscillator sequence (demonstrating superenergy) both exhibiting superoscillations in time.   In the last case, the wavefunction superoscillates everywhere in time in a suitable limit.  Our conclusions are given in Sec.~\ref{sec:conc}.

\section{Superoscillations as a weak value of momentum} \label{sec1}
We begin our analysis by revisiting what a superoscillation is and consider another way to think about it.
Consider plane waves in one dimension as eigenvectors of the momentum operator, ${\hat p}$, ${\hat p} | k\ra = \hbar k |k\ra$, with an unnormalized position representation of $\phi_k(x) = e^{i k x}$. We apply periodic boundary conditions on the region $(-\pi, \pi)$ and restrict the wavenumbers, bounded between $(k_{{\rm min}}, k_{{\rm max}})$.  A new state of the form $\psi(x)  = \sum_j c_j e^{i k_j x}$, can be constructed, 
where $j$ labels the permitted wavenumbers.  This is by definition a band-limited function.  We now define the {\it local wavenumber}
\be
k(x) = {\rm Im} \frac{d}{dx} \ln \psi(x).
\label{local-wn}
\ee
If the state $\psi(x)$ is the eigenstate of momentum, then $k(x) = k$.  However, instead if we choose a general superposition as above, it is possible that $k(x)$ can exceed the band limit $[k_{{\rm min}}, k_{{\rm max}}]$. When $k(x)$ exceeds this range, it is called a {\it superoscillation}.
While this is true for quantum states, it is equally true for any function, viewed from the perspective of Fourier analysis.
Let us take, for example, the commonly found superoscillation function \cite{berry2019roadmap}
\begin{eqnarray}
f(x) &=& \left[ \cos\frac{x}{N} + i a \sin\frac{ x}{N}\right]^N  \nonumber \\ &=& \sum_n C_n e^{i k_n x},
\end{eqnarray}
where $C_n$ are the Fourier coefficients.
The wavenumbers of the superposition are bounded as $k_n \in [-1, 1]$, while the 
the local wavenumber is given by
\be
k(x) = \frac{ \sin(x/N) + a \cos(x/N)}{\cos(x/N) + a \sin(x/N)}.
\label{locwn}
\ee
In the vicinity around $x=0$, $k(x \approx 0) = a$, which can be much larger than 1.

The local wavenumber $k(x)$ can also be seen as a {\it weak value} \cite{aharonov1988result,dressel2014colloquium} of momentum,
\be
\hbar k(x) = {\rm Re} \frac{ \la x | {\hat p} |\psi\ra}{ \la x |\psi\ra},
\ee
once we note that the position representation of the momentum operator is given by ${\hat p} = -i \hbar \partial_x$.  Consequently, the local wavenumber may be viewed as a weak value of momentum, where the pre-selection is on state $|\psi\ra$ (a band-limited superposition of eigenstates of momentum) and the postselection is on position $x$ (a projection on the particle's position). This quantity may also be interpreted as the Bohmian momentum for the initial state, which we can now interpret operationally as the
average momentum conditioned on the subsequent measurement of a particular $x$ in the ideal limit of no measurement disturbance \cite{leavens2005weak,wiseman2007grounding,kocsis2011observing,dressel2015weak}.

If instead of the real part of the weak value, we take the imaginary part, then when the function exceeds the band limit, the function is defined to exhibit {\it supergrowth} - where the local rate of growth of decay of a function exceeds the band limit.  The imaginary part of the weak value is related to the `osmotic velocity' \cite{nelson1966derivation,hiley2012weak,dressel2012significance}.  This supergrowth effect leads to another approach to realize superresolution in optical physics \cite{jordan2020superresolution,karmakar2023supergrowth,karmakar2023experimental}.

Notice there is a sum rule - if we weight the local wavenumber (momentum) with the probability of the postselection, and integrate over the result $x$ we find
\begin{eqnarray}
\int dx |\psi(x)|^2 \hbar k(x) &=& 
 \int dx \psi(x)^\ast  \la x | {\hat p} | \psi\ra  \\ \nonumber
 &=&  \int dx \la \psi | x\ra \la x | {\hat p} | \psi\ra \\ &=& \la \psi | {\hat p} | \psi\ra, \nonumber
\end{eqnarray}
which is the expectation value of the momentum operator in state $|\psi\ra$, as in the weak value case.  Here we used the completeness of the position states.

\section{Generalization to any observable}\label{sec:gen}
The above analysis suggests how to generalize the notion of superoscillation to any observable.  Let us define the eigensystem of the operator $\hat {\cal O}$ (we take to be a  Hermitian observable for simplicity), as $
{\hat {\cal O}} | \phi_l \ra  = \lambda_l | \phi_l \ra$,
where $\lambda_l$ are the eigenvalues and $|\phi_l\ra$ are the eigenstates.  Let us choose to form a new state $| \psi\ra$ using only the eigenstates of ${\hat {\cal O}}$ corresponding to eigenvalues such that $\lambda_{{\rm min}} \le \lambda_j \le \lambda_{{\rm max}}$.  That is, we bound the considered eigenvalues to a range - the analog of a band limit.  We consider a state of the form $| \psi\ra = \sum_j c_j |\phi_j\ra$, where $c_j$ are complex coefficients, and the sum $j$ is bounded as described above.    We define the local super-observable function as the weak value
\be
{\tilde O}(x) = \frac{ \la x | {\hat {\cal O}} |\psi\ra}{ \la x |\psi\ra}. \label{o}
\ee
Taking the real part typically corresponds to the value read off from weak measurement experiments \cite{dressel2014colloquium}, but the imaginary part also has significance in the measurement disturbance \cite{dressel2012significance}.
Here the postselection state is the position, in correspondence to the superoscillation case, but other postselection states can also be considered, as is customary in weak values, such as the momentum eigenstates, or another complete basis.  Notice the sum rule of the previous section generalizes to $\int dx |\psi(x)|^2 {\tilde O}(x) =  \la \psi | {\hat {\cal O}} |\psi\ra$, which gives an interpretation of the weak value as a conditioned average. We call the function ${\tilde O}(x)$ the local superobservable function, because there are certain positions $x$, where this function can exceed the eigenvalue range  
$\lambda_{{\rm min}} \le \lambda_j \le \lambda_{{\rm max}}$.  The superobservable function may be interpreted as the value we can assign to the observable at position $x$, given a state $|\psi\ra$.

\section{Illustrative examples} \label{sec-ilex}
Let us illustrate the superobservable function with some examples.  
Quantum observables such as the angular momentum of a particle on a ring, or the energy levels of a particle in a box map directly on the superoscillation phenomena, so we consider here two examples outside that category - total angular momentum, and the energy of a massive particle in a potential.  

\subsection{Total Angular momentum} \label{sec-am}
Let us begin with (rescaled) total angular momentum, defined with the operator ${\bf {\hat L}}^2/\hbar^2 = - \partial_\theta^2 - \cot \theta \partial_\theta - \csc^2 \theta \partial^2_\phi$.
Here $(\theta, \phi)$ are the polar and azimuthal angles.  The eigenstates of this operator $|l, m\ra$ have eigenvalues $l(l+1)$ with a degeneracy $2l+1$, and have a coordinate representation as the spherical harmonics.  According to our prescription above, the rescaled super-angular momentum is given by ${\tilde L}^2(\theta, \phi)$ by
\be
{\tilde L}^2(\theta, \phi)/\hbar^2 = -\frac{\partial_\theta^2 \psi - \cot \theta\, \partial_\theta \psi - \csc^2 \theta\, \partial^2_\phi \psi}{\psi(\theta, \phi)},
\ee
where $\psi(\theta, \phi)$ is a angle-space wavefunction composed only with eigenstates of $l, m$ that are bounded to a fixed range.  Using just the $m$ variables gives the same physics as superoscillations, so we consider an example with $|\psi\ra \propto |l=0, m=0\ra + c |l=1, m=0\ra$, where $c$ is an arbitrary constant.  In the case $m=0$, the spherical harmonics reduce to the Legendre polynomials with the argument $\cos \theta$. In this case, we find the result ${\tilde L}^2(\theta)/\hbar^2 = 2 c \cos \theta / (1+ c \cos \theta)$.  This function takes on negative values for the range $\theta \in [\pi/2, \pi]$ for $0< c  < 1$, whereas the eigenvalues of the states are $0$ and $2$.  Thus, we see super-behavior in this simple example. For the case $c=1$, a divergence appears at $\theta = \pi$ which persists for $c>1$ - this results from the zero in the wavefunction, which is a generic feature of this sort of super-behavior. - the same as the weak value divergence when the pre- and post-selection states' overlap is zero \cite{aharonov1988result}.

\subsection{Energy}
Consider the Hamiltonian of the system as the observable in question.  We take the canonical form ${\hat H} = {\hat p}^2/(2m) + V({\hat x})$ for a particle of mass $m$ moving in a one dimensional potential $V$ for simplicity, which can have an unbounded spectrum.  The {\it local energy}, ${\tilde E}(x)$, is defined from (\ref{o}) to be 
\be
{\tilde E}(x) = -\frac{\hbar^2}{2m} \frac{1}{\psi(x)} \frac{d^2 \psi(x)}{dx^2} + V(x). \label{loc-ener}
\ee
This basis-dependent result has the simple interpretation as the sum of the potential energy and a kinetic energy term given in terms of the curvature of the wavefunction. 
If $\psi(x)$ is an energy eigenstate $|E_j\ra$ of the system, then by the energy-space Schr\"odinger equation, we obtain simply $E_j$, the correct energy eigenvalue which is the case for all positions $x$.  This definition of the local energy has the appealing form of a rewritten time-independent Schr\"odinger equation, giving an interpretation to the local energy of any wavefunction $\psi(x)$. We now consider a state given by a superposition of energy eigenstates, $|\psi \ra = \sum_j c_j |E_j\ra$,
where the maximum energy is given by $E_{{\rm max}}$. Simple examples can be constructed, similarly to the previous one, where the local energy can exceed this range, and even generally diverge when the wavefunction hits a zero.

\section{Energy out of nothing?}
\label{energy-nothing}
It has been previously shown that in the superoscillation case, a limit may be considered where the range of consideration also extends to the whole domain of the function, so the range of superoscillatory behavior also extends everywhere in this singular limit, despite the function remaining band limited at any finite value of the sequence \cite{aharonov2011some,aharonov2021new}.  We can obtain an energetic analog of this effect by considering a sequence of potentials $V_N(x)$, indexed by a positive integer $N$.  We wish to sum up energy eigenstates with arbitrary coefficients that have a maximum value of $E_{\rm max}$.  As more eigenstates are allowed in the sum, we change the index $N$, such that in each term of the sequence, the maximum energy is still $E_{\rm max}$.  In this way, we can consider a formal limit of $N\rightarrow \infty$ and seek to extend the range of super-behavior.  In the case below, we can even let $E_{\rm max}$ decrease with $N$ and still obtain a convergent result, as we will now show.

Let us illustrate this strategy with the quantum harmonic oscillator.  The Hamiltonian is given by
${\hat H}_N = \frac{{\hat p}^2}{2m} + \frac{1}{2}m \omega_N^2 {\hat x}^2$.
The energy eigenvalues are given by $E^{(N)}_n = \hbar \omega_N (n + 1/2)$. Let us fix the maximum energy we are allowing to be some constant energy $E_{{\rm max}} = \hbar \omega_N(N+1/2)$, so we allow only the quantum index $n$ to extend to $N$ defined by this relation.  This indicates we should scale the sequence of frequencies to be $\omega_N = E_{{\rm max}}/(\hbar(N+1/2)) \approx \omega_0/N$ for large $N$, where $\omega_0$ is related to $E_{{\rm max}}$ as above.
We recall the energy eigenfunctions of the harmonic oscillator are given by
\be
\psi^{(N)}_n(x) = A_n \exp\left(-\frac{m \omega_N x^2}{2 \hbar}\right) H_n\left(\sqrt{\frac{m\omega_N}{\hbar}} x\right),
\ee
where $A_n$ is the normalization constant 
and $H_n(z)$ is the $n$th Hermite polynomial.  These states are also indexed by the label $N$.

\begin{figure}
    \includegraphics[width=7cm]{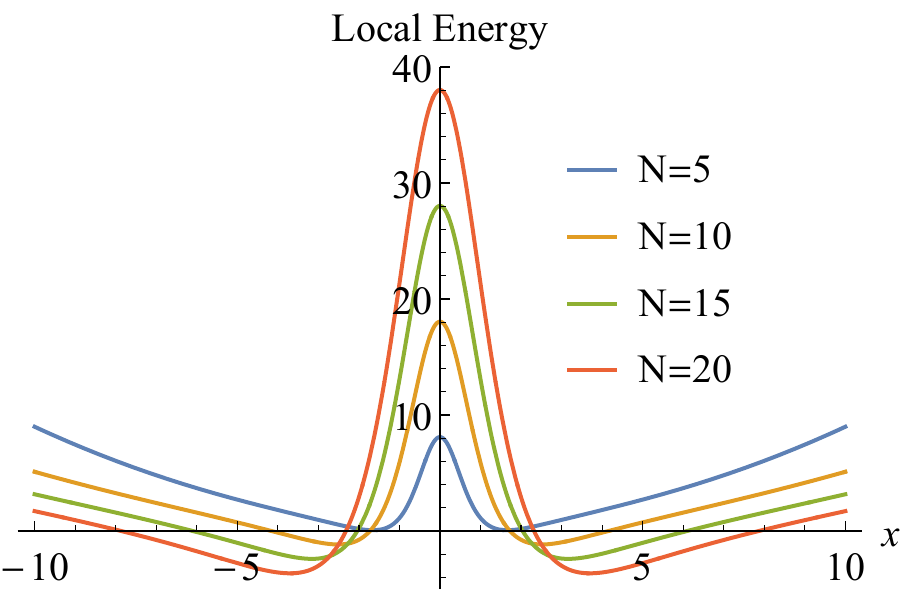}
    \caption{Plot of the real part of the scaled local energy (\ref{loc-ener}) for the scaling $\omega_N = \omega_0/N$.  Values exceeding 1 are in the superenergy range.  As $N$ increases, both the value of the superenergy increases, as well as the region of the superenergy behavior. We choose $g= 0.5$, $m \omega_0/\hbar =1$.
    }
    \label{fig:super}
\end{figure}

We now consider a state constructed as
\be
h_N(x) = \sum_{n=0}^N c^{(N)}_n \psi_n^{(N)}(x), \label{h-const}
\ee
where we let the complex coefficients (also indexed by $N$) take the form
\be
c^{(N)}_n = \frac{1}{A_n} \binom{N}{n} i^n H_{N-n}(g).
\ee
Here $g$ is an arbitrary parameter.  We can sum the series using the identity \cite{wolf}
\be
\sum_{k=0}^N  \binom{N}{k} i^k H_{N-k}(a)H_k(b) = 2^N (a+ib)^N, \label{identity}
\ee
to find the unnormalized state is given by
\begin{eqnarray}
h_N(x) &=& 2^N \exp\left(-\frac{m \omega_N x^2}{2 \hbar}\right) \left[g + i \sqrt{\frac{m\omega_N}{\hbar}} x\right]^N
\label{finiten}
\\
&=& (2 g)^N \exp\left(-\frac{m \omega_N x^2}{2 \hbar}\right) \left[ 1 + \frac{i}{g}
\sqrt{\frac{m\omega_N}{\hbar}} x \right]^N. \nonumber
\end{eqnarray}
In the limit where $N$ is getting large, $\omega_N =\omega_0/N$ becomes increasingly small, so we can approximate,
\be
h_N \approx (2g)^N \exp\left(-\frac{m \omega_0 x^2}{2N \hbar}\right) \exp\left( \frac{i}{g}
\sqrt{\frac{m\omega_0}{\hbar}} \sqrt{N} x \right). \label{hn}
\ee
In the range $|x| < \sqrt{N} g \sqrt{\hbar/m\omega_0}$, we have a local superenergy that {\it grows} with $N$ as $
E_S = \hbar \omega_0 N/(2 g^2)$, corresponding to a superwavenumber of $k_S \propto \sqrt{N}/g$.  Thus, both the value and range grow as $N$ is increased.  Plots of the (real part of) the local energy are shown in Fig.~\ref{fig:super} as $N$ increases.

\begin{figure}[t!]
\centering
       \begin{subfigure}[t]{.3\textwidth}
        \includegraphics[width=\textwidth]{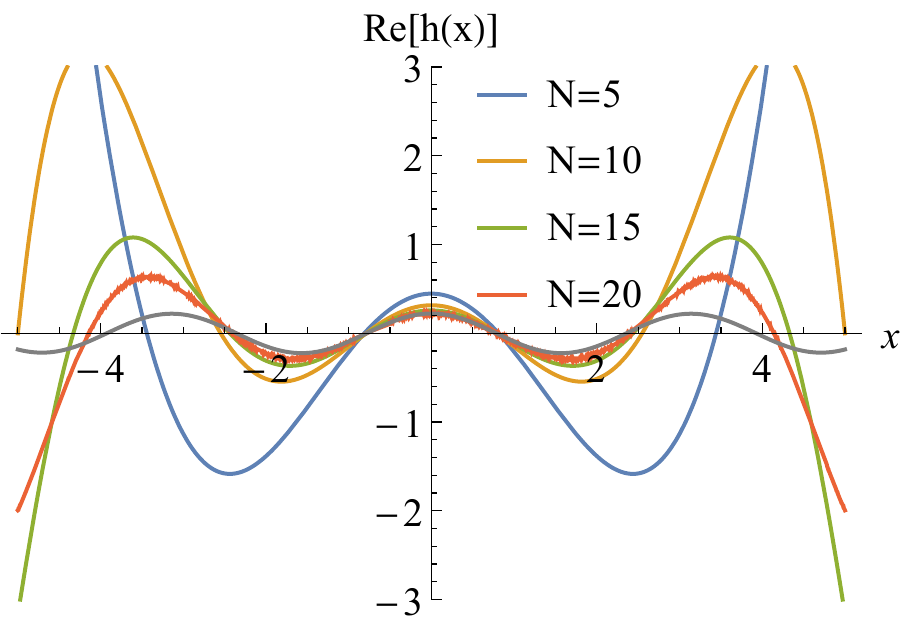}
         \label{fig:rh}
     \end{subfigure}
     \begin{subfigure}[t]{.3\textwidth}
     \centering
         \includegraphics[width=\textwidth]{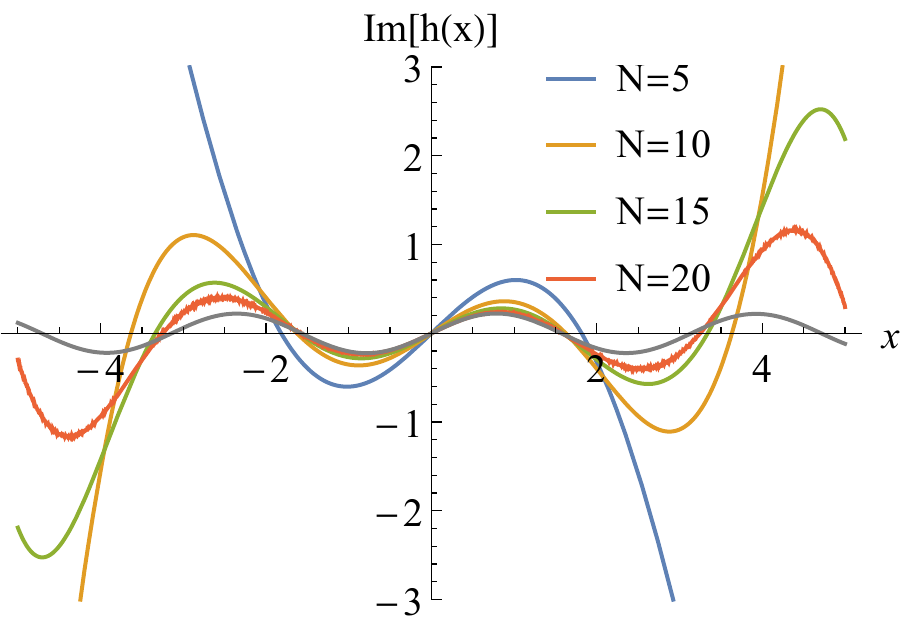}
         \label{fig:ih}
     \end{subfigure}
     \caption{Real (top) and imaginary (bottom) parts of $h_N(x)$ as $N$ is increased.  The limiting form is given in grey for both plots.  Here we take $g=0.5$.}\label{fig:rih}
   \end{figure}

It is both physically and mathematically interesting to choose a different scaling of the frequency of the oscillator with respect to the $N$ index: ${\omega}_N' = \omega_0/N^2$. 
In this case, the maximum energy of the component states decreases as ${E}_n' = \hbar \omega_0 (n + 1/2)/N^2$,
where we keep only $n$ values up to $N$, so the energy maximum decreases inversely with $N$. This scaling choice helps to prove convergence, as well as corresponds to finding the limit of an infinite number of states all with asymptotically zero energy, combining to give a state of seemingly finite energy, a point that we will discuss further below.
The limiting state (\ref{hn}) becomes
\be
{\tilde h}(x) =  \lim_{N\rightarrow \infty} \left(\frac{m \omega_0}{2\pi \hbar N^2}\right)^{1/4} \exp\left(-\frac{m \omega_0 x^2}{2 N^2\hbar} + i \frac{x}{g} \sqrt{\frac{m \omega_0}{\hbar}} \right), \label{hntilde}
\ee
where we renormalize the state, resulting in a Gaussian regularized plane wave with wavenumber $k_0 = \sqrt{m \omega_0/\hbar} /g$.  The width of the Gaussian regularization is $N \sqrt{\hbar/(m  \omega_0)}$, which diverges as $N \rightarrow \infty$, leaving a finite energy plane wave.  Convergence to this solution (plotted in grey) is shown in the Fig.~\ref{fig:rih} as $N$ is increased for the real and imaginary part of ${\tilde h}_N(x)$.

\begin{figure}[tb]
    \centering
    \includegraphics[width=6cm]{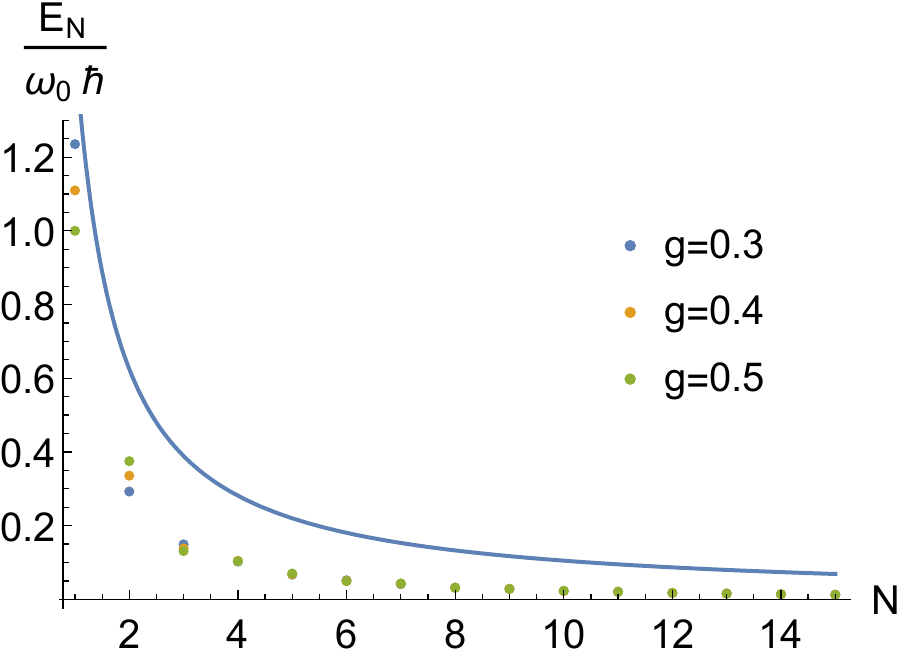}
    \caption{The scaled energy expectation value of the normalized states are plotted versus $N$ for different values of $g$.  The solid curve is the bound of the upper energy eigenvalue in the superposition, $(N+1/2)/N^2$.}
    \label{fig:ebound}
\end{figure}

\section{Quantum mimicry}\label{sec-qmim}
The above situation is paradoxical:  We have proven the convergence of the sequence of states to a finite energy plane wave, when every finite element of the sequence has an energy that decreases as $~1/N$.  That is, for any fixed value of $x$, the limit as $N\rightarrow \infty$ is given by Eq.~(\ref{hntilde}). 

Further insight into this situation can be had by computing the expected energy in the state in two different ways.  The first way is to make the eigenfunction decomposition as done in Eq.~(\ref{h-const}).  From this point of view, the normalized expected energy of the state is given by
\be
E_N = \frac{ \sum_{n=1}^N |c^{(N)}_n|^2 (n + 1/2) \hbar \omega_N}{ \sum_{n=1}^N |c^{(N)}_n|^2}.
\ee
The decay to zero of the energy expectation value of the sequence of states is illustrated in Fig.~\ref{fig:ebound}, and is bounded from above by $E_N/(\hbar \omega_0) \le (N+ 1/2)/N^2$.

\begin{figure}[tb]
\centering
    \includegraphics[width=8cm]{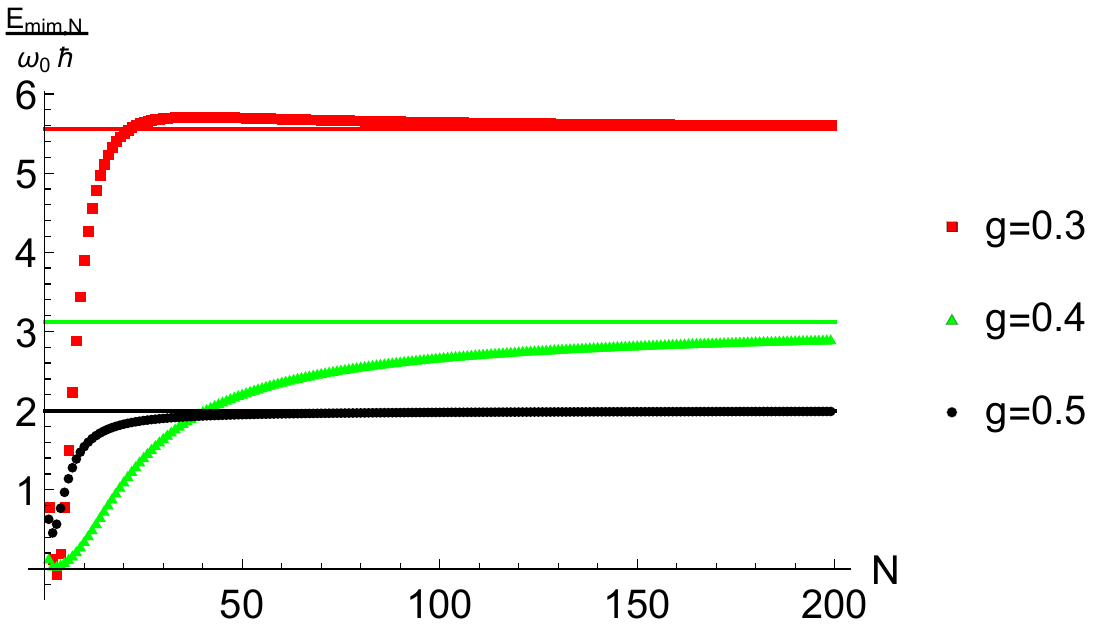}
    \caption{The scaled local energy $E_{mim,N}$ (\ref{mimic}) is plotted as a function of $N$ for different values of $g$.  The range is taken to be $(-2, 2)$ in this plot, where $m \omega_0/\hbar =1$.  Convergence to the local energy $\hbar \omega_0 /2g^2$, shown as horizontal lines, is seen for large enough $N$.}
    \label{fig:locconverge}
\end{figure}

On the other hand, we can also calculate the expected energy in a finite interval $(-L, L)$ of space as
\be
E_{mim,N} = \frac{\int_{-L}^{L} dx h_N(x) {\hat H}_N h_N(x)}  { \int_{-L}^{L} dx |h_N(x)|^2 }. \label{mimic}
\ee
When we take $L \rightarrow \infty$ for fixed $N$ the two calculations will converge; however, if instead we set $L = L_N \le g \sqrt{ \hbar N/ m \omega_0}$, then in the limit where $N \rightarrow \infty$, (\ref{mimic}) gives the local energy of ${\tilde h}$, $\hbar \omega_0/(2 g^2)$.  This effect is shown in Fig.~\ref{fig:locconverge}.  There we plot Eq.~(\ref{mimic}) as function of $N$ for different values of $g$.  The interval is fixed between (-2,2) in units where $m \omega_0/\hbar =1$, which satisfies the above condition on $L_N$ for the chosen parameter values at the upper end of the $N$ range. 

We note that Eq.~(\ref{mimic}) has the natural interpretation of the conditional energy of a particle when it is postselected in the spatial region $(-L, L)$. 
The denominator represents the postselection probability of finding the particle in the spatial region. The postselection suppresses the undesired growth of the function outside the interval $(-L, L)$, which restores the low energy behavior of the entire function.  This viewpoint is reminiscent of the `red to gamma' claim of Aharonov and colleagues \cite{aharonov1990can}. This claim can now be more dramatically put as the energetic `zero to hero':  a coherent superposition of a diverging number of asymptotically zero energy states conspire to make up a finite energy state in a given spatial region.
It has been showed elsewhere that such a postselection resulting in a much larger energy than can be seen from the component states comprising the superposition came from the preparation step and requires a quantum reference frame \cite{aharonov2023conservation}.  This other source of energy resolves the energy conservation puzzle.

We have nevertheless shown something remarkable:  for any finite $N$ in the sequence, the Schr\"odinger equation is satisfied, and the constructed function using negligible amounts of energy is able to {\it mimic} a high energy state for as large a spatial region as desired.  The cost for this mimicry is that outside that region the function must blow up, such that the total energy of the function is actually very small when normalized properly.  On the scale of whole function, the super-energy portion is exponentially small in magnitude, similar to the superoscillation case.  Nevertheless, the size of the spatial region of superbehavior can become arbitrary large.

\section{Superoscillations in time}  \label{sec-sotime}
In this section, we show that superbehavior of the quantum mechanical observable leads to superoscillation in the complimentary generating variable.  Considering the time dynamics of the state gives additional insight into the super behavior of the dynamics.  Allowing the local energy $E(x,t)$ defined in Eq.~(\ref{loc-ener}) to be time-dependent through the pre-selected state, $\psi(x, t)$, we have the following result by the time-dependent Schr\"odinger equation,
\be
E(x, t) = i \hbar \frac{1}{ \psi(x, t)} \frac{\partial}{\partial t} \psi(x, t) = i \hbar \frac{\partial}{\partial t} \ln \psi(x, t). \label{loc-energy2}
\ee
Thus the time-dependent local energy is equivalent to the weak value of the operator
$i \hbar \partial_t$, preselected on the state $|\psi(t)\ra$ and postselected on position $x$,
\be
E(x, t) = \frac{\la x |i \hbar \partial_t|\psi(t)\ra }{\la x | \psi(t)\ra}.
\ee
Here we note an internal consistency with the discussion and results of Sec.~\ref{sec1}.  We note that a superposition of time-dependent energy eigenstates $|\phi_j\ra$ is of the form,
\be
|\psi(t)\ra = \sum_j c_j e^{-i E_j t/\hbar }|\phi_j\ra.
\ee
Consequently, for a single energy-eigenstate, we have $i \hbar \partial_t|\phi_j(t)\ra = E_j |\phi_j(t)\ra$.
Therefore it is natural to ascribe the real part of the weak value (\ref{loc-energy2}) as the local (time and space-dependent) energy.  We note that this is a similar to the concept of the local wavenumber (\ref{local-wn}), so the superposed wavefunction locally behaves in time like $\psi(x, t) \sim \exp(-i t E(x, t)/\hbar) \psi(x, 0)$.

When the local energy exceeds the energy eigenvalue bounds of the superposition, it immediately follows that the wavefunction {\it superoscillates in time}. That is, the local oscillation frequency in time exceeds energy eigenvalue bounds of the superposition. The mathematics of this observation is exactly the same as the usual superoscillation case, but applied to time rather than space.

This is a specific case of a more general phenomenon.  Consider the generating function of an observable $\hat {\cal O}$, given by $Z(\chi) =\la \phi | \exp (i \chi 
{\hat {\cal O}})| \psi \ra/\la \phi| \psi\ra $, so successive derivatives of $Z$ with respect to $\chi$ generate all weak valued moments of the observable. We now consider the preselected state to be a superposition of eigenstates of $\cal \hat O$ corresponding to eigenvalues bounded between $\lambda_{\rm min}$ and $\lambda_{max}$.  In the weak generating function, we can approximate 
\be
Z(\chi) \approx  \exp (i \chi 
{\tilde {\cal O}}_w),
\ee
where ${\tilde {\cal O}}_w$ is the corresponding weak value.  Therefore, when the weak value exceeds its eigenstate bounds for specific choices of pre and post-selected states, the generating function superoscillates as a function of the generating variable $\chi$. That is, the oscillation frequency of $Z$ with respect to $\chi$ exceeds that of the maximum eigenvalue or is lower than that of the lowest eigenvalue.  The previous example corresponds to setting $\la \phi|$ to a position eigenstate, the operator to be the Hamiltonian, and the generating variable to be the time.

\subsection{Application to total angular momentum}

We can apply the previous examples of superphenomena to illustrate the effect of connecting the super angular momentum or Hamiltonian to the superoscillation in time. We consider the quantum mechanics of a three dimensional rotor of mass $m$ and fixed length $a$.  The Hamiltonian is then given by ${\hat H} = {\bf \hat L}^2/(2 m a^2)$.  We consider the same example in Sec.~\ref{sec-am}, and exampling the time-dependence of the local Hamiltonian, controlled here by the squared total angular momentum.  In this case, the dependence of the post-selected state in time is given by
\be
\psi(\theta, t) \approx \psi(\theta, 0) \exp\left( - \frac{i t \hbar}{2ma^2} \frac{2c \cos\theta}{(1+c \cos \theta)} \right).  
\ee
The pre-selection of a superposition of $l=0,1$ implies that the post-selected wavefunction will 
superoscillate in time when ever there is superbehavior in the total angular momentum, when the oscillation frequency exceeds the range $[0, \hbar/ma^2]$.

\subsection{Application to superenergy in the harmonic oscillator sequence}
Let us now examine the time-dependence of the state (\ref{finiten}).
We consider the unnormalized state
\be
h_N(x) = e^{-z^2/2} \sum_{n=0}^N  \binom{N}{n} i^n H_{N-n}(g) H_n(z),
\ee
where $z = \sqrt{\frac{m \omega_0}{\hbar}} \frac{x}{N}$.
Time dynamics can be inserted by multiplying each eigenstate inside the sum by $\exp( - i E_n t/\hbar)$, where $E_n = \hbar \omega_N (n+1/2)$ and $t$ is the time.  Thus, the time-evolved state is
\be
h_N(x, t) = e^{-z^2/2 - i \omega_N t/2} \sum_{n=0}^N  \binom{N}{n} e^{i \varphi n} H_{N-n}(g) H_n(z),\label{time-evo}
\ee
where $\varphi = \pi/2 - \omega_N t$.  We see immediately that the state is in fact periodic as is typical of harmonic oscillator solution, when $\omega_N t$ is an integer multiple of $2\pi$ and the function is the complex conjugate of itself when $\omega_N t$ is an odd multiple of $\pi$ (up to an overall phase). Recalling our second scaling ${\tilde \omega}_N = \omega_0/N^2$, this cycle period diverges as $N \rightarrow \infty$.   

To make further progress, we recall the Hermite polynomials are generated by the function
\be
H_n(x) =\left. \frac{d^n}{du^n} e^{2 x u - u^2}\right\vert_{u=0}.
\ee
Using two copies of the generating function, we find
\begin{widetext}

\be
h_N(x, t) = e^{-z^2/2 - i \omega_N t/2} \left.\sum_{n=0}^N  \binom{N}{n} e^{i \varphi n} \frac{\partial^{N-n}}{\partial u^{N-n}} \frac{\partial^n}{\partial v^n}  e^{2 g u - u^2 + 2 z v - v^2}\right\vert_{u=v=0}. \nonumber
\ee
The binomial sum can now be carried out to find
\be
h_N(x, t) = \left.e^{-z^2/2 - i \omega_N t/2}  \left[ \frac{\partial}{\partial u} + e^{i\varphi} \frac{\partial}{\partial v} \right]^N e^{2 g u - u^2 + 2 z v - v^2}\right\vert_{u=v=0}. 
\ee
When $e^{i\varphi} = i$ the result (\ref{identity}) is straightforward to prove by induction.  It may also be seen by defining a complex variable $\zeta = u + i v$, so the quadratic term in the exponent is $ - {\bar \zeta} \zeta$, where $\bar \zeta$ denotes the complex conjugate of $\zeta$.  The differential operator in the preceding equation then becomes $2^N \partial^N/\partial {\bar \zeta}^N$ so
\begin{eqnarray}
2^n \frac{\partial^n}{\partial {\bar \zeta}^n} \exp \left[ \zeta(g- i z) + {\bar \zeta }(g + i z) - \zeta {\bar \zeta}\right]&=&2^n \frac{\partial^{n-1}}{\partial {\bar \zeta}^{n-1}} (g  + i z - \zeta) \exp\left[ \zeta(g- i z) + {\bar \zeta }(g + i z) - \zeta {\bar \zeta}\right],
\\
&=& 2^n  (g  + i z - \zeta)^n \exp\left[ \zeta(g- i z) + {\bar \zeta }(g + i z) - \zeta {\bar \zeta}\right]. \label{complexder}
\end{eqnarray}
In the calculation above, we used the fact that  $\partial \zeta/\partial {\bar \zeta}=0$.  Setting $\zeta=0$ proves the quoted result (\ref{identity}).  

For a fixed time $t$, the limit $N \rightarrow \infty$ seemingly removes the time-dependence in the above equations.  This is because $\varphi = t \omega_0/N^2$, while the highest derivative power is of order $N$, so the time dependence is at best $~e^{t/N}$, which is eliminated as $N\rightarrow \infty$.  However, this estimate is incorrect, because there is a superoscillation effect in time, so the sums of these seemingly negligible terms constructively add to survive in the large $N$ limit, as we will now prove.  This effect is a reflection of the sums of eigenstates with energies that limit to zero giving a finite energy state.   

We consider the first order correction in time to $h_N(x, t)$.  Taking $\omega_N = \omega_0/N^2$, we have the result
\be
h_N(x, t) \approx \left.e^{-z^2/2 - i \omega_N t/2}  \left[ \frac{\partial}{ \partial u} + i \frac{\partial}{\partial v} + \frac{\omega_0 t}{N^2} \frac{\partial}{\partial v} \right]^N e^{2 g u - u^2 + 2 z v - v^2}\right\vert_{u=v=0}. 
\ee
We now make a binomial approximation to leading order, treating the term $\omega_0 t/N^2$ as small to find

\be
h_N(x, t) \approx \left.e^{-z^2/2 - i \omega_N t/2} \left( 2^N (g + i z)^N +   \frac{\omega_0 t}{N} \frac{\partial}{\partial v} \left[ \frac{\partial}{ \partial u} + i \frac{\partial}{\partial v}\right]^{N-1} e^{2 g u - u^2 + 2 z v - v^2}\right\vert_{u=v=0}\right). 
\ee
The second term can be calculated by using result (\ref{complexder}) with $n=N-1$ and taking a final $v$ derivative before setting $u, v=0$.  The result is
\be
h_N(z, t) \approx e^{-z^2/2 - i \omega_N t/2} \left( 2^N (g + i z)^N +   \frac{\omega_0 t}{N} 2^{N-1} \left(2 z (g+iz)^{N-1} - i (N-1) (g+i z)^{N-2} \right)\right). 
\ee
Importantly, one of the terms has an $(N-1)$ which together with the $N$ from the binomial expansion compensates the $1/N^2$ term, leaving a finite first order in time correction as $N$ is taken to infinity.  We can expand this observation to higher orders in the power series expansion in time to find the contribution remaining in the large $N$ limit is given by
\be
h_N(z, t) \approx e^{-z^2/2} \left( 2^N (g + i z)^N \right)
\sum_{n = 0}^{N/2}  \frac{N!}{n! (N-2n)! N^{2n}} \left(\frac{-i \omega_0 t} {2(g+iz)^2}\right)^n.
\ee
\end{widetext}
In the limit $n \ll N$, where the series converges rapidly for $ t < \hbar/ E(x, t) \approx 2 g^2/\omega_0$ (given we are in the super-energy range of $|z|<g$), we notice that the sum is close to an exponential series, because the
combination
\be
\ln(N!/((N-2n)! N^{2n})) \approx -2 n^2/N.
\ee
Here we used the Stirling approximation, and therefore we can drop this part of the prefactor of the power series for small orders. 

\begin{figure}[t!]
\centering
       \begin{subfigure}[t]{.45\textwidth}
        \includegraphics[width=\textwidth]{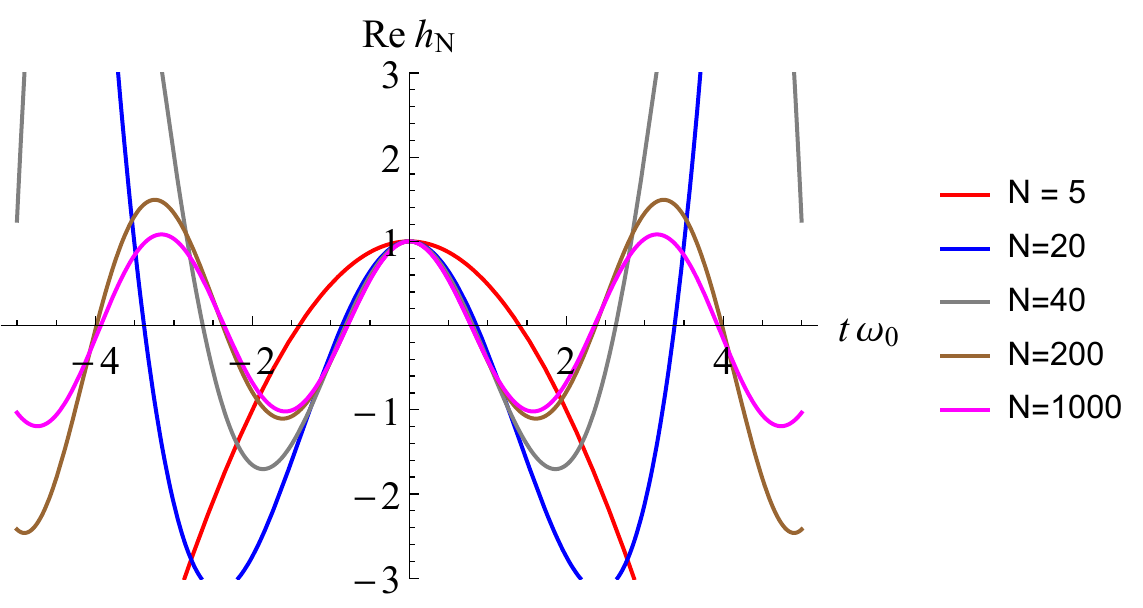}
         \label{fig:sore}
     \end{subfigure}
     \begin{subfigure}[t]{.45\textwidth}
     \centering \includegraphics[width=\textwidth]{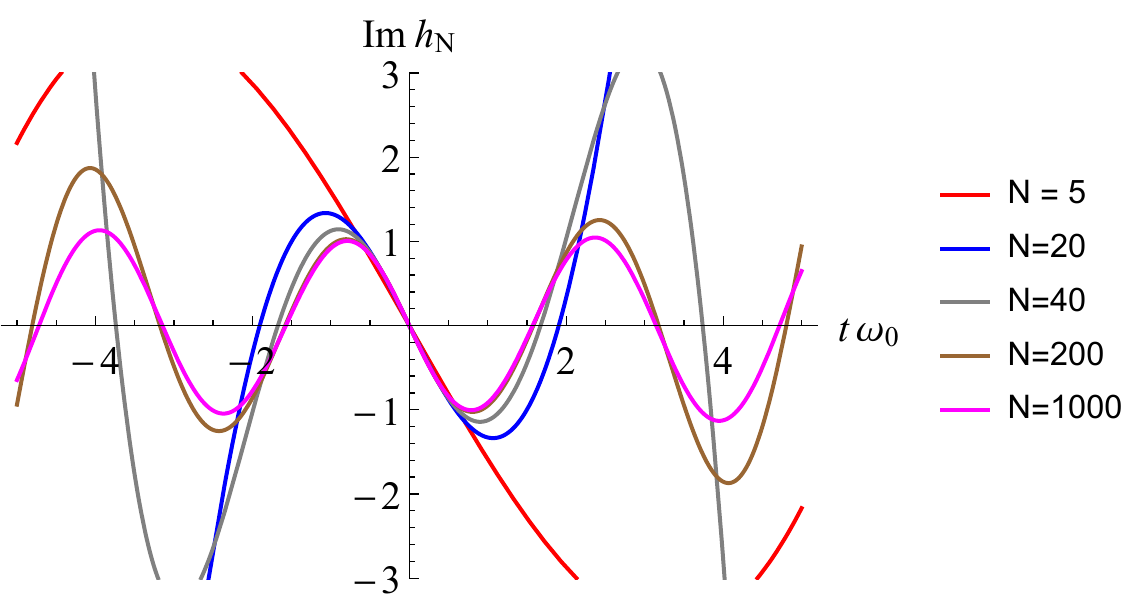}
         \label{fig:soim}
     \end{subfigure}
     \caption{Real (top) and imaginary (bottom) parts of $h_N(0, t)$ (\ref{time-evo}) is plotted versus time as $N$ is increased, for $x=0$.  The solution converges to $\exp(-2 i \omega_0 t)$ for the choice $g=0.5$.}\label{fig:sointime}
   \end{figure}

We therefore find the approximation for $|z|<g$, or $|x| < N g \sqrt{\hbar/m \omega_0}$, and $|t| <2 g^2/\omega_0$ to be
\be
h_N(z, t) \approx (2 g)^N \exp\left(-z^2/2 + i z N/g - i \omega_0 t/(2 g^2)\right),
\ee
showing the time-dependence of state (\ref{hntilde}). Here we display the superenergy near the origin of $E_S = \hbar \omega_0 /(2 g^2)$ that was derived independently in Sec.~\ref{energy-nothing}. 
Restoring the definition $z=\sqrt{m \omega_0/\hbar} x/N$, we see this solution describes a plane wave in time and space with a phase velocity of $v_p = \sqrt{\hbar \omega_0/m}/(2g)$.
Convergence to the superoscillation in time is shown in Fig.~\ref{fig:sointime} as the index $N$ is increased, up to $N=1000$.   Therefore, we also have superoscillations in time near $t=0$, with the superenergy, as was showed in general in the beginning of this section.  Before concluding, we note the numerical simulations show the region in time where there is good convergence to the superoscillation in time expands as $N$ increases, so we expect a good approximation is had for time much less than 
 $2 N g^2/\omega_0$, similar to the range of superenergy in space, which produces a superoscillation everywhere in time as well.


\section{Conclusions}
\label{sec:conc}
By returning to the origins of superoscillations, we recognized their form as a weak value of a momentum operator, which in turn motivated a general definition of super-behavior for any quantum operator:  A pre-selected function is prepared that is a superposition of eigenstates of said observable with a bounded spectrum of eigenvalues.  The operator is then postselected on the particle's position.  When the resulting weak value exceeds the eigenspectrum of the operator at a given position, super-behavior occurs.  It is natural to further define generalizations such as postselection on momentum or energy eigenstates, however, we have focused in this article on position, as the natural extension of the superoscillation case.  Examples of local super-angular momentum and energy were discussed.  Notably, we found a sequence of potentials and corresponding states such that the mathematical limit produced a finite energy state everywhere on the real line, constructed from states whose asymptotic energy went to zero in the limit.  This construction demonstrates that finite energy states can be mimicked by states of asymptotically zero energy, for as large a spatial region as desired.  We also saw that superenergy behavior in space implied superoscillations in time, with a frequency given by the superenergy, divided by the reduced Planck's constant.  This effect may be useful for super-resolution of features of quantum potentials.

The connection to generalized superoscillations may be reestablished by pointing out that although we have focused on quantum mechanical effects, our findings may be applied quite generally to spectral theory in mathematics.  More specifically, differential operators have a spectral decomposition in terms of eigenvalues and eigenfunctions in the theory of differential equations.   By superimposing the eigenfunctions of such operators with a bounded spectral band, solutions may be constructed that mimic a eigenfunction of the same operator corresponding to an eigenvalue outside that spectral band over an arbitrarily large range, and that converges everywhere in a suitable mathematical limit.

{\it Acknowledgments.---}
This work was supported by the AFOSR grant \#FA9550-21-1-0322 and the Bill Hannon Foundation.  We thank Sandu Popescu for helpful discussions.


\begin{thebibliography}{29}%
\makeatletter
\providecommand \@ifxundefined [1]{%
 \@ifx{#1\undefined}
}%
\providecommand \@ifnum [1]{%
 \ifnum #1\expandafter \@firstoftwo
 \else \expandafter \@secondoftwo
 \fi
}%
\providecommand \@ifx [1]{%
 \ifx #1\expandafter \@firstoftwo
 \else \expandafter \@secondoftwo
 \fi
}%
\providecommand \natexlab [1]{#1}%
\providecommand \enquote  [1]{``#1''}%
\providecommand \bibnamefont  [1]{#1}%
\providecommand \bibfnamefont [1]{#1}%
\providecommand \citenamefont [1]{#1}%
\providecommand \href@noop [0]{\@secondoftwo}%
\providecommand \href [0]{\begingroup \@sanitize@url \@href}%
\providecommand \@href[1]{\@@startlink{#1}\@@href}%
\providecommand \@@href[1]{\endgroup#1\@@endlink}%
\providecommand \@sanitize@url [0]{\catcode `\\12\catcode `\$12\catcode
  `\&12\catcode `\#12\catcode `\^12\catcode `\_12\catcode `\%12\relax}%
\providecommand \@@startlink[1]{}%
\providecommand \@@endlink[0]{}%
\providecommand \url  [0]{\begingroup\@sanitize@url \@url }%
\providecommand \@url [1]{\endgroup\@href {#1}{\urlprefix }}%
\providecommand \urlprefix  [0]{URL }%
\providecommand \Eprint [0]{\href }%
\providecommand \doibase [0]{https://doi.org/}%
\providecommand \selectlanguage [0]{\@gobble}%
\providecommand \bibinfo  [0]{\@secondoftwo}%
\providecommand \bibfield  [0]{\@secondoftwo}%
\providecommand \translation [1]{[#1]}%
\providecommand \BibitemOpen [0]{}%
\providecommand \bibitemStop [0]{}%
\providecommand \bibitemNoStop [0]{.\EOS\space}%
\providecommand \EOS [0]{\spacefactor3000\relax}%
\providecommand \BibitemShut  [1]{\csname bibitem#1\endcsname}%
\let\auto@bib@innerbib\@empty
\bibitem [{\citenamefont {Berry}\ and\ \citenamefont
  {Popescu}(2006)}]{berry2006evolution}%
  \BibitemOpen
  \bibfield  {author} {\bibinfo {author} {\bibfnamefont {M.}~\bibnamefont
  {Berry}}\ and\ \bibinfo {author} {\bibfnamefont {S.}~\bibnamefont
  {Popescu}},\ }\bibfield  {title} {\bibinfo {title} {Evolution of quantum
  superoscillations and optical superresolution without evanescent waves},\
  }\href@noop {} {\bibfield  {journal} {\bibinfo  {journal} {Journal of Physics
  A: Mathematical and General}\ }\textbf {\bibinfo {volume} {39}},\ \bibinfo
  {pages} {6965} (\bibinfo {year} {2006})}\BibitemShut {NoStop}%
\bibitem [{\citenamefont {Aharonov}\ \emph {et~al.}(2017)\citenamefont
  {Aharonov}, \citenamefont {Colombo}, \citenamefont {Sabadini}, \citenamefont
  {Struppa},\ and\ \citenamefont {Tollaksen}}]{aharonov2017mathematics}%
  \BibitemOpen
  \bibfield  {author} {\bibinfo {author} {\bibfnamefont {Y.}~\bibnamefont
  {Aharonov}}, \bibinfo {author} {\bibfnamefont {F.}~\bibnamefont {Colombo}},
  \bibinfo {author} {\bibfnamefont {I.}~\bibnamefont {Sabadini}}, \bibinfo
  {author} {\bibfnamefont {D.}~\bibnamefont {Struppa}},\ and\ \bibinfo {author}
  {\bibfnamefont {J.}~\bibnamefont {Tollaksen}},\ }\href@noop {} {\emph
  {\bibinfo {title} {The mathematics of superoscillations}}},\ Vol.\ \bibinfo
  {volume} {247}\ (\bibinfo  {publisher} {American Mathematical Society},\
  \bibinfo {year} {2017})\BibitemShut {NoStop}%
\bibitem [{\citenamefont {Berry}\ \emph {et~al.}(2019)\citenamefont {Berry},
  \citenamefont {Zheludev}, \citenamefont {Aharonov}, \citenamefont {Colombo},
  \citenamefont {Sabadini}, \citenamefont {Struppa}, \citenamefont {Tollaksen},
  \citenamefont {Rogers}, \citenamefont {Qin}, \citenamefont {Hong} \emph
  {et~al.}}]{berry2019roadmap}%
  \BibitemOpen
  \bibfield  {author} {\bibinfo {author} {\bibfnamefont {M.}~\bibnamefont
  {Berry}}, \bibinfo {author} {\bibfnamefont {N.}~\bibnamefont {Zheludev}},
  \bibinfo {author} {\bibfnamefont {Y.}~\bibnamefont {Aharonov}}, \bibinfo
  {author} {\bibfnamefont {F.}~\bibnamefont {Colombo}}, \bibinfo {author}
  {\bibfnamefont {I.}~\bibnamefont {Sabadini}}, \bibinfo {author}
  {\bibfnamefont {D.~C.}\ \bibnamefont {Struppa}}, \bibinfo {author}
  {\bibfnamefont {J.}~\bibnamefont {Tollaksen}}, \bibinfo {author}
  {\bibfnamefont {E.~T.}\ \bibnamefont {Rogers}}, \bibinfo {author}
  {\bibfnamefont {F.}~\bibnamefont {Qin}}, \bibinfo {author} {\bibfnamefont
  {M.}~\bibnamefont {Hong}}, \emph {et~al.},\ }\bibfield  {title} {\bibinfo
  {title} {Roadmap on superoscillations},\ }\href@noop {} {\bibfield  {journal}
  {\bibinfo  {journal} {Journal of Optics}\ }\textbf {\bibinfo {volume} {21}},\
  \bibinfo {pages} {053002} (\bibinfo {year} {2019})}\BibitemShut {NoStop}%
\bibitem [{\citenamefont {Kempf}(2018)}]{kempf2018four}%
  \BibitemOpen
  \bibfield  {author} {\bibinfo {author} {\bibfnamefont {A.}~\bibnamefont
  {Kempf}},\ }\bibfield  {title} {\bibinfo {title} {Four aspects of
  superoscillations},\ }\href@noop {} {\bibfield  {journal} {\bibinfo
  {journal} {Quantum Studies: Mathematics and Foundations}\ }\textbf {\bibinfo
  {volume} {5}},\ \bibinfo {pages} {477} (\bibinfo {year} {2018})}\BibitemShut
  {NoStop}%
\bibitem [{\citenamefont {Huang}\ \emph {et~al.}(2007)\citenamefont {Huang},
  \citenamefont {Chen}, \citenamefont {de~Abajo},\ and\ \citenamefont
  {Zheludev}}]{huang2007optical}%
  \BibitemOpen
  \bibfield  {author} {\bibinfo {author} {\bibfnamefont {F.~M.}\ \bibnamefont
  {Huang}}, \bibinfo {author} {\bibfnamefont {Y.}~\bibnamefont {Chen}},
  \bibinfo {author} {\bibfnamefont {F.~J.~G.}\ \bibnamefont {de~Abajo}},\ and\
  \bibinfo {author} {\bibfnamefont {N.~I.}\ \bibnamefont {Zheludev}},\
  }\bibfield  {title} {\bibinfo {title} {Optical super-resolution through
  super-oscillations},\ }\href@noop {} {\bibfield  {journal} {\bibinfo
  {journal} {Journal of Optics A: Pure and Applied Optics}\ }\textbf {\bibinfo
  {volume} {9}},\ \bibinfo {pages} {S285} (\bibinfo {year} {2007})}\BibitemShut
  {NoStop}%
\bibitem [{\citenamefont {Huang}\ and\ \citenamefont
  {Zheludev}(2009)}]{huang2009super}%
  \BibitemOpen
  \bibfield  {author} {\bibinfo {author} {\bibfnamefont {F.~M.}\ \bibnamefont
  {Huang}}\ and\ \bibinfo {author} {\bibfnamefont {N.~I.}\ \bibnamefont
  {Zheludev}},\ }\bibfield  {title} {\bibinfo {title} {Super-resolution without
  evanescent waves},\ }\href@noop {} {\bibfield  {journal} {\bibinfo  {journal}
  {Nano letters}\ }\textbf {\bibinfo {volume} {9}},\ \bibinfo {pages} {1249}
  (\bibinfo {year} {2009})}\BibitemShut {NoStop}%
\bibitem [{\citenamefont {Rogers}\ \emph {et~al.}(2012)\citenamefont {Rogers},
  \citenamefont {Lindberg}, \citenamefont {Roy}, \citenamefont {Savo},
  \citenamefont {Chad}, \citenamefont {Dennis},\ and\ \citenamefont
  {Zheludev}}]{rogers2012super}%
  \BibitemOpen
  \bibfield  {author} {\bibinfo {author} {\bibfnamefont {E.~T.}\ \bibnamefont
  {Rogers}}, \bibinfo {author} {\bibfnamefont {J.}~\bibnamefont {Lindberg}},
  \bibinfo {author} {\bibfnamefont {T.}~\bibnamefont {Roy}}, \bibinfo {author}
  {\bibfnamefont {S.}~\bibnamefont {Savo}}, \bibinfo {author} {\bibfnamefont
  {J.~E.}\ \bibnamefont {Chad}}, \bibinfo {author} {\bibfnamefont {M.~R.}\
  \bibnamefont {Dennis}},\ and\ \bibinfo {author} {\bibfnamefont {N.~I.}\
  \bibnamefont {Zheludev}},\ }\bibfield  {title} {\bibinfo {title} {A
  super-oscillatory lens optical microscope for subwavelength imaging},\
  }\href@noop {} {\bibfield  {journal} {\bibinfo  {journal} {Nature materials}\
  }\textbf {\bibinfo {volume} {11}},\ \bibinfo {pages} {432} (\bibinfo {year}
  {2012})}\BibitemShut {NoStop}%
\bibitem [{\citenamefont {Zheludev}(2008)}]{zheludev2008diffraction}%
  \BibitemOpen
  \bibfield  {author} {\bibinfo {author} {\bibfnamefont {N.~I.}\ \bibnamefont
  {Zheludev}},\ }\bibfield  {title} {\bibinfo {title} {What diffraction
  limit?},\ }\href@noop {} {\bibfield  {journal} {\bibinfo  {journal} {Nature
  materials}\ }\textbf {\bibinfo {volume} {7}},\ \bibinfo {pages} {420}
  (\bibinfo {year} {2008})}\BibitemShut {NoStop}%
\bibitem [{\citenamefont {Greenfield}\ \emph {et~al.}(2013)\citenamefont
  {Greenfield}, \citenamefont {Schley}, \citenamefont {Hurwitz}, \citenamefont
  {Nemirovsky}, \citenamefont {Makris},\ and\ \citenamefont
  {Segev}}]{greenfield2013experimental}%
  \BibitemOpen
  \bibfield  {author} {\bibinfo {author} {\bibfnamefont {E.}~\bibnamefont
  {Greenfield}}, \bibinfo {author} {\bibfnamefont {R.}~\bibnamefont {Schley}},
  \bibinfo {author} {\bibfnamefont {I.}~\bibnamefont {Hurwitz}}, \bibinfo
  {author} {\bibfnamefont {J.}~\bibnamefont {Nemirovsky}}, \bibinfo {author}
  {\bibfnamefont {K.~G.}\ \bibnamefont {Makris}},\ and\ \bibinfo {author}
  {\bibfnamefont {M.}~\bibnamefont {Segev}},\ }\bibfield  {title} {\bibinfo
  {title} {Experimental generation of arbitrarily shaped diffractionless
  superoscillatory optical beams},\ }\href@noop {} {\bibfield  {journal}
  {\bibinfo  {journal} {Optics Express}\ }\textbf {\bibinfo {volume} {21}},\
  \bibinfo {pages} {13425} (\bibinfo {year} {2013})}\BibitemShut {NoStop}%
\bibitem [{\citenamefont {Makris}\ and\ \citenamefont
  {Psaltis}(2011)}]{makris2011superoscillatory}%
  \BibitemOpen
  \bibfield  {author} {\bibinfo {author} {\bibfnamefont {K.~G.}\ \bibnamefont
  {Makris}}\ and\ \bibinfo {author} {\bibfnamefont {D.}~\bibnamefont
  {Psaltis}},\ }\bibfield  {title} {\bibinfo {title} {Superoscillatory
  diffraction-free beams},\ }\href@noop {} {\bibfield  {journal} {\bibinfo
  {journal} {Optics letters}\ }\textbf {\bibinfo {volume} {36}},\ \bibinfo
  {pages} {4335} (\bibinfo {year} {2011})}\BibitemShut {NoStop}%
\bibitem [{\citenamefont {Aharonov}\ \emph {et~al.}(1990)\citenamefont
  {Aharonov}, \citenamefont {Popescu},\ and\ \citenamefont
  {Rohrlich}}]{aharonov1990can}%
  \BibitemOpen
  \bibfield  {author} {\bibinfo {author} {\bibfnamefont {Y.}~\bibnamefont
  {Aharonov}}, \bibinfo {author} {\bibfnamefont {S.}~\bibnamefont {Popescu}},\
  and\ \bibinfo {author} {\bibfnamefont {D.}~\bibnamefont {Rohrlich}},\
  }\bibfield  {title} {\bibinfo {title} {How can an infra-red photon behave as
  a gamma ray},\ }\href@noop {} {\bibfield  {journal} {\bibinfo  {journal}
  {Tel-Aviv University Preprint TAUP}\ ,\ \bibinfo {pages} {1847}} (\bibinfo
  {year} {1990})}\BibitemShut {NoStop}%
\bibitem [{\citenamefont {Aharonov}\ and\ \citenamefont
  {Vaidman}(1990)}]{aharonov1990properties}%
  \BibitemOpen
  \bibfield  {author} {\bibinfo {author} {\bibfnamefont {Y.}~\bibnamefont
  {Aharonov}}\ and\ \bibinfo {author} {\bibfnamefont {L.}~\bibnamefont
  {Vaidman}},\ }\bibfield  {title} {\bibinfo {title} {Properties of a quantum
  system during the time interval between two measurements},\ }\href@noop {}
  {\bibfield  {journal} {\bibinfo  {journal} {Physical Review A}\ }\textbf
  {\bibinfo {volume} {41}},\ \bibinfo {pages} {11} (\bibinfo {year}
  {1990})}\BibitemShut {NoStop}%
\bibitem [{\citenamefont {Berry}(1994)}]{berry1994faster}%
  \BibitemOpen
  \bibfield  {author} {\bibinfo {author} {\bibfnamefont {M.}~\bibnamefont
  {Berry}},\ }\href@noop {} {\bibinfo {title} {Faster than fourier quantum
  coherence and reality; in celebration of the 60th birthday of yakir aharonov
  ed js anandan and jl safko}} (\bibinfo {year} {1994})\BibitemShut {NoStop}%
\bibitem [{\citenamefont {Aharonov}\ \emph {et~al.}(2011)\citenamefont
  {Aharonov}, \citenamefont {Colombo}, \citenamefont {Sabadini}, \citenamefont
  {Struppa},\ and\ \citenamefont {Tollaksen}}]{aharonov2011some}%
  \BibitemOpen
  \bibfield  {author} {\bibinfo {author} {\bibfnamefont {Y.}~\bibnamefont
  {Aharonov}}, \bibinfo {author} {\bibfnamefont {F.}~\bibnamefont {Colombo}},
  \bibinfo {author} {\bibfnamefont {I.}~\bibnamefont {Sabadini}}, \bibinfo
  {author} {\bibfnamefont {D.}~\bibnamefont {Struppa}},\ and\ \bibinfo {author}
  {\bibfnamefont {J.}~\bibnamefont {Tollaksen}},\ }\bibfield  {title} {\bibinfo
  {title} {Some mathematical properties of superoscillations},\ }\href@noop {}
  {\bibfield  {journal} {\bibinfo  {journal} {Journal of Physics A:
  Mathematical and Theoretical}\ }\textbf {\bibinfo {volume} {44}},\ \bibinfo
  {pages} {365304} (\bibinfo {year} {2011})}\BibitemShut {NoStop}%
\bibitem [{\citenamefont {Aharonov}\ \emph {et~al.}(2021)\citenamefont
  {Aharonov}, \citenamefont {Colombo}, \citenamefont {Sabadini}, \citenamefont
  {Shushi}, \citenamefont {Struppa},\ and\ \citenamefont
  {Tollaksen}}]{aharonov2021new}%
  \BibitemOpen
  \bibfield  {author} {\bibinfo {author} {\bibfnamefont {Y.}~\bibnamefont
  {Aharonov}}, \bibinfo {author} {\bibfnamefont {F.}~\bibnamefont {Colombo}},
  \bibinfo {author} {\bibfnamefont {I.}~\bibnamefont {Sabadini}}, \bibinfo
  {author} {\bibfnamefont {T.}~\bibnamefont {Shushi}}, \bibinfo {author}
  {\bibfnamefont {D.~C.}\ \bibnamefont {Struppa}},\ and\ \bibinfo {author}
  {\bibfnamefont {J.}~\bibnamefont {Tollaksen}},\ }\bibfield  {title} {\bibinfo
  {title} {A new method to generate superoscillating functions and
  supershifts},\ }\href@noop {} {\bibfield  {journal} {\bibinfo  {journal}
  {Proceedings of the Royal Society A}\ }\textbf {\bibinfo {volume} {477}},\
  \bibinfo {pages} {20210020} (\bibinfo {year} {2021})}\BibitemShut {NoStop}%
\bibitem [{\citenamefont {Aharonov}\ \emph {et~al.}(1988)\citenamefont
  {Aharonov}, \citenamefont {Albert},\ and\ \citenamefont
  {Vaidman}}]{aharonov1988result}%
  \BibitemOpen
  \bibfield  {author} {\bibinfo {author} {\bibfnamefont {Y.}~\bibnamefont
  {Aharonov}}, \bibinfo {author} {\bibfnamefont {D.~Z.}\ \bibnamefont
  {Albert}},\ and\ \bibinfo {author} {\bibfnamefont {L.}~\bibnamefont
  {Vaidman}},\ }\bibfield  {title} {\bibinfo {title} {How the result of a
  measurement of a component of the spin of a spin-1/2 particle can turn out to
  be 100},\ }\href@noop {} {\bibfield  {journal} {\bibinfo  {journal} {Physical
  review letters}\ }\textbf {\bibinfo {volume} {60}},\ \bibinfo {pages} {1351}
  (\bibinfo {year} {1988})}\BibitemShut {NoStop}%
\bibitem [{\citenamefont {Dressel}\ \emph {et~al.}(2014)\citenamefont
  {Dressel}, \citenamefont {Malik}, \citenamefont {Miatto}, \citenamefont
  {Jordan},\ and\ \citenamefont {Boyd}}]{dressel2014colloquium}%
  \BibitemOpen
  \bibfield  {author} {\bibinfo {author} {\bibfnamefont {J.}~\bibnamefont
  {Dressel}}, \bibinfo {author} {\bibfnamefont {M.}~\bibnamefont {Malik}},
  \bibinfo {author} {\bibfnamefont {F.~M.}\ \bibnamefont {Miatto}}, \bibinfo
  {author} {\bibfnamefont {A.~N.}\ \bibnamefont {Jordan}},\ and\ \bibinfo
  {author} {\bibfnamefont {R.~W.}\ \bibnamefont {Boyd}},\ }\bibfield  {title}
  {\bibinfo {title} {Colloquium: Understanding quantum weak values: Basics and
  applications},\ }\href@noop {} {\bibfield  {journal} {\bibinfo  {journal}
  {Reviews of Modern Physics}\ }\textbf {\bibinfo {volume} {86}},\ \bibinfo
  {pages} {307} (\bibinfo {year} {2014})}\BibitemShut {NoStop}%
\bibitem [{\citenamefont {Leavens}(2005)}]{leavens2005weak}%
  \BibitemOpen
  \bibfield  {author} {\bibinfo {author} {\bibfnamefont {C.}~\bibnamefont
  {Leavens}},\ }\bibfield  {title} {\bibinfo {title} {Weak measurements from
  the point of view of bohmian mechanics},\ }\href@noop {} {\bibfield
  {journal} {\bibinfo  {journal} {Foundations of Physics}\ }\textbf {\bibinfo
  {volume} {35}},\ \bibinfo {pages} {469} (\bibinfo {year} {2005})}\BibitemShut
  {NoStop}%
\bibitem [{\citenamefont {Wiseman}(2007)}]{wiseman2007grounding}%
  \BibitemOpen
  \bibfield  {author} {\bibinfo {author} {\bibfnamefont {H.}~\bibnamefont
  {Wiseman}},\ }\bibfield  {title} {\bibinfo {title} {Grounding bohmian
  mechanics in weak values and bayesianism},\ }\href@noop {} {\bibfield
  {journal} {\bibinfo  {journal} {New Journal of Physics}\ }\textbf {\bibinfo
  {volume} {9}},\ \bibinfo {pages} {165} (\bibinfo {year} {2007})}\BibitemShut
  {NoStop}%
\bibitem [{\citenamefont {Kocsis}\ \emph {et~al.}(2011)\citenamefont {Kocsis},
  \citenamefont {Braverman}, \citenamefont {Ravets}, \citenamefont {Stevens},
  \citenamefont {Mirin}, \citenamefont {Shalm},\ and\ \citenamefont
  {Steinberg}}]{kocsis2011observing}%
  \BibitemOpen
  \bibfield  {author} {\bibinfo {author} {\bibfnamefont {S.}~\bibnamefont
  {Kocsis}}, \bibinfo {author} {\bibfnamefont {B.}~\bibnamefont {Braverman}},
  \bibinfo {author} {\bibfnamefont {S.}~\bibnamefont {Ravets}}, \bibinfo
  {author} {\bibfnamefont {M.~J.}\ \bibnamefont {Stevens}}, \bibinfo {author}
  {\bibfnamefont {R.~P.}\ \bibnamefont {Mirin}}, \bibinfo {author}
  {\bibfnamefont {L.~K.}\ \bibnamefont {Shalm}},\ and\ \bibinfo {author}
  {\bibfnamefont {A.~M.}\ \bibnamefont {Steinberg}},\ }\bibfield  {title}
  {\bibinfo {title} {Observing the average trajectories of single photons in a
  two-slit interferometer},\ }\href@noop {} {\bibfield  {journal} {\bibinfo
  {journal} {Science}\ }\textbf {\bibinfo {volume} {332}},\ \bibinfo {pages}
  {1170} (\bibinfo {year} {2011})}\BibitemShut {NoStop}%
\bibitem [{\citenamefont {Dressel}(2015)}]{dressel2015weak}%
  \BibitemOpen
  \bibfield  {author} {\bibinfo {author} {\bibfnamefont {J.}~\bibnamefont
  {Dressel}},\ }\bibfield  {title} {\bibinfo {title} {Weak values as
  interference phenomena},\ }\href@noop {} {\bibfield  {journal} {\bibinfo
  {journal} {Physical Review A}\ }\textbf {\bibinfo {volume} {91}},\ \bibinfo
  {pages} {032116} (\bibinfo {year} {2015})}\BibitemShut {NoStop}%
\bibitem [{\citenamefont {Nelson}(1966)}]{nelson1966derivation}%
  \BibitemOpen
  \bibfield  {author} {\bibinfo {author} {\bibfnamefont {E.}~\bibnamefont
  {Nelson}},\ }\bibfield  {title} {\bibinfo {title} {Derivation of the
  schr{\"o}dinger equation from newtonian mechanics},\ }\href@noop {}
  {\bibfield  {journal} {\bibinfo  {journal} {Physical review}\ }\textbf
  {\bibinfo {volume} {150}},\ \bibinfo {pages} {1079} (\bibinfo {year}
  {1966})}\BibitemShut {NoStop}%
\bibitem [{\citenamefont {Hiley}(2012)}]{hiley2012weak}%
  \BibitemOpen
  \bibfield  {author} {\bibinfo {author} {\bibfnamefont {B.}~\bibnamefont
  {Hiley}},\ }\bibfield  {title} {\bibinfo {title} {Weak values: Approach
  through the clifford and moyal algebras},\ }in\ \href@noop {} {\emph
  {\bibinfo {booktitle} {Journal of Physics: Conference Series}}},\ Vol.\
  \bibinfo {volume} {361}\ (\bibinfo {organization} {IOP Publishing},\ \bibinfo
  {year} {2012})\ p.\ \bibinfo {pages} {012014}\BibitemShut {NoStop}%
\bibitem [{\citenamefont {Dressel}\ and\ \citenamefont
  {Jordan}(2012)}]{dressel2012significance}%
  \BibitemOpen
  \bibfield  {author} {\bibinfo {author} {\bibfnamefont {J.}~\bibnamefont
  {Dressel}}\ and\ \bibinfo {author} {\bibfnamefont {A.~N.}\ \bibnamefont
  {Jordan}},\ }\bibfield  {title} {\bibinfo {title} {Significance of the
  imaginary part of the weak value},\ }\href@noop {} {\bibfield  {journal}
  {\bibinfo  {journal} {Physical Review A}\ }\textbf {\bibinfo {volume} {85}},\
  \bibinfo {pages} {012107} (\bibinfo {year} {2012})}\BibitemShut {NoStop}%
\bibitem [{\citenamefont {Jordan}(2020)}]{jordan2020superresolution}%
  \BibitemOpen
  \bibfield  {author} {\bibinfo {author} {\bibfnamefont {A.~N.}\ \bibnamefont
  {Jordan}},\ }\bibfield  {title} {\bibinfo {title} {Superresolution using
  supergrowth and intensity contrast imaging},\ }\href@noop {} {\bibfield
  {journal} {\bibinfo  {journal} {Quantum Studies: Mathematics and
  Foundations}\ }\textbf {\bibinfo {volume} {7}},\ \bibinfo {pages} {285}
  (\bibinfo {year} {2020})}\BibitemShut {NoStop}%
\bibitem [{\citenamefont {Karmakar}\ \emph
  {et~al.}(2023{\natexlab{a}})\citenamefont {Karmakar}, \citenamefont
  {Chakraborty}, \citenamefont {Vamivakas},\ and\ \citenamefont
  {Jordan}}]{karmakar2023supergrowth}%
  \BibitemOpen
  \bibfield  {author} {\bibinfo {author} {\bibfnamefont {T.}~\bibnamefont
  {Karmakar}}, \bibinfo {author} {\bibfnamefont {A.}~\bibnamefont
  {Chakraborty}}, \bibinfo {author} {\bibfnamefont {A.~N.}\ \bibnamefont
  {Vamivakas}},\ and\ \bibinfo {author} {\bibfnamefont {A.~N.}\ \bibnamefont
  {Jordan}},\ }\bibfield  {title} {\bibinfo {title} {Supergrowth and
  sub-wavelength object imaging},\ }\href@noop {} {\bibfield  {journal}
  {\bibinfo  {journal} {Optics Express}\ }\textbf {\bibinfo {volume} {31}},\
  \bibinfo {pages} {37174} (\bibinfo {year} {2023}{\natexlab{a}})}\BibitemShut
  {NoStop}%
\bibitem [{\citenamefont {Karmakar}\ \emph
  {et~al.}(2023{\natexlab{b}})\citenamefont {Karmakar}, \citenamefont {Wadood},
  \citenamefont {Jordan}, \citenamefont {Vamivakas} \emph
  {et~al.}}]{karmakar2023experimental}%
  \BibitemOpen
  \bibfield  {author} {\bibinfo {author} {\bibfnamefont {T.}~\bibnamefont
  {Karmakar}}, \bibinfo {author} {\bibfnamefont {S.}~\bibnamefont {Wadood}},
  \bibinfo {author} {\bibfnamefont {A.~N.}\ \bibnamefont {Jordan}}, \bibinfo
  {author} {\bibfnamefont {A.~N.}\ \bibnamefont {Vamivakas}}, \emph {et~al.},\
  }\bibfield  {title} {\bibinfo {title} {Experimental realization of
  supergrowing fields},\ }\href@noop {} {\bibfield  {journal} {\bibinfo
  {journal} {arXiv preprint arXiv:2309.00016}\ } (\bibinfo {year}
  {2023}{\natexlab{b}})}\BibitemShut {NoStop}%
\bibitem [{wol()}]{wolf}%
  \BibitemOpen
  \href@noop {} {\bibinfo {title} {Wolfram functions, howpublished =
  {\url{http://functions.wolfram.com/05.01.23.0002.01}}, note = {Accessed:
  2024-01-23}}}\BibitemShut {NoStop}%
\bibitem [{\citenamefont {Aharonov}\ \emph {et~al.}(2023)\citenamefont
  {Aharonov}, \citenamefont {Popescu},\ and\ \citenamefont
  {Rohrlich}}]{aharonov2023conservation}%
  \BibitemOpen
  \bibfield  {author} {\bibinfo {author} {\bibfnamefont {Y.}~\bibnamefont
  {Aharonov}}, \bibinfo {author} {\bibfnamefont {S.}~\bibnamefont {Popescu}},\
  and\ \bibinfo {author} {\bibfnamefont {D.}~\bibnamefont {Rohrlich}},\
  }\bibfield  {title} {\bibinfo {title} {Conservation laws and the foundations
  of quantum mechanics},\ }\href@noop {} {\bibfield  {journal} {\bibinfo
  {journal} {Proceedings of the National Academy of Sciences}\ }\textbf
  {\bibinfo {volume} {120}},\ \bibinfo {pages} {e2220810120} (\bibinfo {year}
  {2023})}\BibitemShut {NoStop}%
\end{thebibliography}
\end{document}